\documentclass[smallextended]{svjour3}       
\smartqed  
\usepackage{xcolor}
\usepackage{amsmath}
\usepackage{upgreek}
\usepackage[title]{appendix}
\usepackage{epsfig}
\usepackage{graphics}
\usepackage{psfrag}
\usepackage{epstopdf}
\usepackage{amssymb}
\usepackage{relsize}
\usepackage{bm}
\usepackage{mathrsfs}
\usepackage[center]{caption}
\usepackage[none]{hyphenat}
\usepackage{setspace}
\doublespace
\journalname{EURASIP Journal on WCN}

\begin{document}

\newcommand{\dd}{\mathop{}\mathopen{}\mathrm{d}}

\title{Cross-layer distributed power control: A repeated games formulation to improve the sum energy-efficiency}



\titlerunning{Cross-layer distributed power control: A RG formulation to improve the sum EE}        

\author{Mariem Mhiri,          
        Vineeth S. Varma,      
				Karim Cheikhrouhou,    
				Samson Lasaulce       and
				Abdelaziz Samet
}

\authorrunning{M. Mhiri, V. S. Varma, K. Cheikhrouhou, S. Lasaulce and A. Samet} 

\institute{M. Mhiri \at
             SERCOM Laboratory, Tunisia Polytechnic School, P.B. 743-2078, La Marsa, Tunisia \\
             \email{mariem.mhiri@gmail.com}          
          \and
           V. S. Varma \at
              Singapore University of Technology and Design, Singapore \\
							\email{vineethsvarma@gmail.com}
					\and
					 K. Cheikhrouhou \at
              SERCOM Laboratory, Tunisia Polytechnic School, P.B. 743-2078, La Marsa, Tunisia \\
              \email{karim.cheikhrouhou@gmail.com}           
          \and
					 S. Lasaulce \at
              L2S - CNRS - SUPELEC, 91192 Gif-sur-Yvette, University of Paris-Sud, France \\
              \email{samson.lasaulce@lss.supelec.fr}    
				  \and
					 A. Samet \at
              EMT Centre - INRS University, Ouest Montr\'{e}al (Qu\'{e}bec), H5A 1K6, Canada \\
              \email{samet@emt.inrs.ca}    						
}

\date{Received: date / Accepted: date}

\maketitle

\begin{abstract}
The main objective of this work is to improve the energy-efficiency (EE) of a multiple access channel (MAC) system, through power control, in a distributed manner. In contrast with many existing works on energy-efficient power control, which ignore the possible presence of a queue at the transmitter, we consider a new generalized cross-layer EE metric. This approach is relevant when the transmitters have a non-zero energy cost even when the radiated power is zero and takes into account the presence of a finite packet buffer and packet arrival at the transmitter. As the Nash equilibrium (NE) is an energy-inefficient solution, the present work aims at overcoming this deficit by improving the global energy-efficiency. Indeed, as the considered system has multiple agencies each with their own interest, the performance metric reflecting the individual interest of each decision maker is the global energy-efficiency defined then as the sum over individual energy-efficiencies. Repeated games (RG) are investigated through the study of two dynamic games (finite RG and discounted RG), whose equilibrium is defined when introducing a new operating point (OP), Pareto-dominating the NE and relying only on individual channel state information (CSI). Accordingly, closed-form expressions of the minimum number of stages of the game for finite RG (FRG) and the maximum discount factor of the discounted RG (DRG) were established. Our contributions consist of improving the system performances in terms of powers and utilities when using the new OP compared to the NE and the Nash bargaining (NB) solution. Moreover, the cross-layer model in the RG formulation leads to achieving a shorter minimum number of stages in the FRG even for higher number of users. In addition, the social welfare (sum of utilities) in the DRG decreases slightly with the cross-layer model when the number of users increases while it is reduced considerably with the Goodman model. Finally, we show that in real systems with random packet arrivals, the cross-layer power control algorithm outperforms the Goodman algorithm.

\end{abstract}

\section{Introduction}

\subsection{Motivation}

The design and management of green wireless networks \cite{lister-09,palicot-2005,greentouch-10} has become increasingly important for modern wireless networks, in particular, to manage operating costs. Futuristic (beyond 5G) cellular networks face the dual challenges of being able to respond to the explosion of data rates and also to manage network energy consumption. Due to the limited spectrum and large number of active users in modern networks, energy-efficient distributed power control is an important issue. Sensor networks, which have multiple sensors sending information to a common receiver with a limited energy, capacity have also recently surged in popularity. Energy minimization in sensor networks has been analysed in many recent works \cite{Yu2006,Bandyopadhyay2003,Cardei2005}.\\
Several of the above described systems have some common features:
\begin{enumerate}
\item Multiple transmitters connected to a common receiver.
\item Lack of centralization or coordination, i.e., a distributed and de-centralized network.
\item Relevance of minimizing energy consumption or maximizing energy-efficiency (EE).
\item Transmitters that have arbitrary data transmission.
\end{enumerate}
These features are present in many modern systems like a sensor network which has multiple sensors with limited energy connected in a distributed manner to a common receiver. These sensors don't always have information to transmit, resulting in sporadic data transmission. Another example would be several mobile devices connected to a hot-spot (via wifi or even Bluetooth). Due to these features of the network, inter-transmitter communication is not possible and the transmitters are independent decision makers. Therefore, implementing frequency or time division multiple access becomes harder and a MAC protocol (with single carrier) is often the preferred or natural method of channel access.

\subsection{Novelty}

In many existing works, both network-centric and user-centric approaches have been studied. In a network-centric approach, the global energy-efficiency (GEE) is defined as the ratio between the system benefit (sum-throughput or sum-rate) over the total cost in terms of consumed power \cite{Zappone2015,Isheden2012}. However, when targeting an efficient solution in an user-centric problem, the GEE becomes not ideal as it has no significance to any of the decision makers. In this case, other metrics are required to reflect the individual interest of each decision maker. Therefore, we redefine the GEE to be the sum over individual energy-efficiencies as a suitable metric of interest \cite{LasaulceTWC2009}.\\
The major novelty of this work is in improving the \textbf{sum of energy-efficiencies} for a communication system with \textbf{all the listed features above}. In such a \textbf{decentralized and distributed network}, as each transmitter operates independently, implementing a frequency division or a time division multiple access is not trivial. Therefore, we are interested in looking at a \textbf{MAC system} where all transmitters operate on the same band. Additionally, \textbf{EE} will be our preferred metric due to its relevance. This metric has been defined in \cite{Goodman2000} as the ratio between the average net data rate and the transmitted power. In \cite{Betz2008,Zappone2013}, the total power consumed by the transmitter was taken into account in the EE expression to design distributed power control which is one of the most well known techniques for improving EE. However, many of the works available on energy-efficient power control consider the EE defined in \cite{Goodman2000} where the possible presence of a queue at the transmitter is ignored. In contrast with the existing works, we consider a new generalized EE based on a cross-layer approach developed recently in \cite{VarmaICC2012,VarmaTVT2014}. This approach is important since it takes into account: 1) a fixed cost in terms of power namely, a cost which does not depend on the radiated power; and 2) \textbf{the presence of a finite packet buffer and sporadic packet arrival} at the transmitter (which corresponds to including the 4th feature mentioned above).
Although providing a more general model, the distributed system in \cite{VarmaTVT2014} may operate at a point which is energy-inefficient. Indeed, the point at which the system operates is a Nash equilibrium (NE) of a certain non-cooperative static game. The present work aims at filling this gap by not only considering a cross-layer approach of energy-efficient power control but also improving the system performance in terms of sum of energy-efficiencies.

\subsection{State of the art}

Nash bargaining (NB) solution in a cooperative game can provide a possible efficient solution concept for the problem of interest as it is Pareto-efficient. However, it generally requires global channel state information (CSI) \cite{MhiriNov2012}. Therefore, we are interested in improving the average performance of the system by considering long-term utilities. We focus then on repeated games (RG) where repetition allows efficient equilibrium points to be implemented. Unlike static games which are played in one shot, RG are a special case of dynamic games which consider a cooperation plan and consist in repeating at each step the same static game and the utilities result from averaging the static game utilities over time \cite{Lasaulce2009}. There are two relevant dynamic RG models: finite (FRG) and discounted (DRG). The FRG is defined when the number of stages during which the players interact is finite. For the DRG model, the discount factor is seen as the stopping probability at each stage \cite{Lasaulce2011}. The power control problem using the classic EE developed by Goodman et al in \cite{Goodman2000} has been solved with RG only in \cite{LeTreust2010} where authors developed an operating point (OP) relying on individual CSI and showed that RG lead to efficient distributed solution. Here, we investigate the power control problem of a MAC system by referring to RG (finite and discounted) where the utility function is based on a cross-layer approach. Accordingly, we contribute to:
\begin{enumerate}
\item determine the closed-form expressions of the minimum number of stages for the FRG and the maximum discount factor for the DRG. These two parameters identify the two considered RG.
\item determine a distributed solution Pareto-dominating the NE and improving the system performances in terms of powers and utilities compared not only to the NE but also to the NB solution even for high number of users.
\item show that the RG formulation when using the new EE and the new OP leads to significant gains in terms of social welfare (sum of utilities of all the users) compared to the NE. 
\item show that the following aspects of the cross-layer model improve considerably the system performances when comparing to the Goodman model even for large number of users:
\begin{itemize}
	\item the minimum number of stages in the cross-layer EE model can always be shorter than the minimum number of stages in the Goodman EE formulation.
	\item the social welfare for the DRG in the cross-layer model decreases slightly when the number of users increases while it decreases considerably in the Goodman model.
	\end{itemize}
	\item show that in real systems with random packet arrivals, the cross-layer power control algorithm outperforms the Goodman algorithm and then the new OP with the cross-layer approach is more efficient.
\end{enumerate}

\subsection{Structure}

This paper is structured as follows. In section \ref{sec:ProblemStatement}, we define the system model under study, introduce the generalized EE metric and define the non-cooperative static game. This is followed (section \ref{sec:NashBargainingSolution}) by the study of the NB solution. In section \ref{sec:RepeatedGamesFormulation}, we introduce the new OP, give the formulation of both RG models (FRG and DRG) and determine the closed-form expressions of the minimum number of stages and the maximum discount factor as well. Numerical results are presented in section \ref{sec:NumericalResults} and finally we draw several concluding remarks.

\section{Problem statement}
\label{sec:ProblemStatement}

\subsection{System model}
\label{sec:SystemModel}

We consider a MAC system composed of $N$ small transmitters communicating with a receiver. The $i^{th}$ transmitter transmits a signal $x_i$ with a power $p_i \in [0, P_{i}^{\max}]$ where $P_{i}^{\max}$ is the maximum transmit power assumed identical for all users ($P_{i}^{\max}=P^{\max}$). The additive noise, which is the same for all users, is an additive white Gaussian noise denoted as $n$ with zero mean and variance $\sigma^{2}$. We assume that the users transmit their data over block fading channels. The channel gain between user $i$ and the receiver is given by $g_{i}$. Thus, the baseband signal received at the receiver is written as:  
\begin{equation}
y = \sum_{\substack{i=1}}^{N} x_{i}|g_{i}|^2 + n. 
\label{1}
\end{equation}
Therefore, the resulting SINR $\gamma_i$ corresponding to the $i^{th}$ transmitter is given by \cite{LeTreust2010,Mhiri2013}:
\begin{equation}
\gamma_{i}(\mathbf{p}) = \displaystyle{\frac{p_{i}|g_{i}|^2}{\sigma^{2}+\sum_{j \neq i} p_{j}|g_{j}|^2 }},
\label{2}
\end{equation}
where $\mathbf{p}=(p_1,p_2,\ldots,p_N)$ defines the power vector of all users and can be written as $\mathbf{p}=(p_i,\mathbf{p_{-i}})$ with $\mathbf{p_{-i}}=(p_1,\ldots,p_{i-1},p_{i+1},\ldots,p_N)$.

The purpose of this work is to determine how each user is going to control its power in an optimum way. Game theory, as a powerful mathematical tool, helps to solve such an optimization problem where the utility function is the EE which is a function of the users powers. Since the system under study has multiple agencies each with individual interest, the sum over individual energy-efficiencies will be considered as the performance metric reflecting the individual interest of each decision maker.

\subsection{Energy-efficiency metric}
\label{sec:EnergyEfficiencyMetric}

The EE is defined in \cite{Goodman2000} as a ratio of the net data rate to the transmit power level and is given by:
\begin{equation} 
\chi_{i}(\mathbf{p})=\frac{Rf(\gamma_{i}(\mathbf{p}))}{p_{i}},
\label{3}
\end{equation}
where $R$ is the transmission rate (in bit/s) while $f : \left[0, +\infty \right) \rightarrow [0,1]$ denotes the efficiency function which is sigmoidal and corresponds to the packet success rate verifying $f(0)=0$ and $\lim\limits_{x \rightarrow +\infty} f(x)=1$. Authors of \cite{Betz2008} were the first to consider a total transmission cost of the type \emph{radiated power} ($p_i$) $+$ \emph{consumed power} ($b$) to design distributed power control strategies for multiple access channels \cite{VarmaICC2012,VarmaTVT2014} as follows:
\begin{equation}
\chi_i(\mathbf{p}) = \frac{Rf(\gamma_i(\mathbf{p}))}{b+p_i}.
\label{eqchi}
\end{equation} 
In \cite{VarmaICC2012,VarmaTVT2014}, a more generalized EE metric has been developed by considering a packet arrival process following a Bernoulli process with a constant probability $q$ and a finite memory buffer of size $K$. The new EE expression is given by:
\begin{equation}
\chi_i(\mathbf{p}) = \displaystyle{\frac{Rq(1-\Phi(\gamma_i(\mathbf{p})))}{b+\displaystyle {\frac{qp_i(1-\Phi(\gamma_i(\mathbf{p})))}{f(\gamma_i(\mathbf{p}))}}}},
\label{eqEEgen}
\end{equation}
where the function $\Phi$ identifies the packet loss due to both bad channel conditions and the finiteness of the packet buffer and is expressed as follows: 
\begin{equation}
\Phi(\gamma_i)=(1-f(\gamma_i))\Pi_K(\gamma_i), 
\label{5}
\end{equation}
where $\Pi_K(\gamma_i)$ is the stationary probability that the buffer is full and is given by:
\begin{equation}
\Pi_K(\gamma_i)=\displaystyle {\frac{\rho^K(\gamma_i)}{1+\rho(\gamma_i)+\ldots+\rho^K(\gamma_i)}},
\label{6}
\end{equation}
with:
\begin{equation}
\rho(\gamma_i)=\displaystyle{\frac{q(1-f(\gamma_i))}{(1-q)f(\gamma_i)}}.
\label{7}
\end{equation}
It is important to highlight that this new generalized EE given by (\ref{eqEEgen}) includes the conventional case of (\ref{eqchi}) when making $q\rightarrow 1$. \\

\subsection{Static cross-layer power control game}
\label{sec:StaticCrossLayerPowerControlGame}

The static cross-layer power control game is a non-cooperative game which can be defined as a strategic form game \cite{Lasaulce2011}.
\begin{definition}
\textit{The game is defined by the ordered triplet $\mathcal{G} = \bigl(\mathcal{N}, (\mathcal{S}_i)_{i \in \mathcal{N}}, \allowbreak (u_i)_{i \in \mathcal{N}} \bigr)$ where $\mathcal{N}$ is the set of players (the $N$ transmitters), $\mathcal{S}_1,\ldots,\mathcal{S}_N$ are the corresponding sets of strategies with $\mathcal{S}_i=[0, P_i^{\max}]$ and $u_1,\ldots,u_N$ are the utility functions given by:}
\begin{equation}
u_i(\mathbf{p}) = \chi_i(\mathbf{p}),
\label{8}
\end{equation}
\end{definition}
\textit{where $\chi_i(\mathbf{p})$ is given by equation (\ref{eqEEgen}).}

In a non-cooperative game, each user (player) seeks to maximize selfishly its individual utility function. The optimum solution results then by setting $\partial u_i/\partial p_i$ to zero as follows:
\begin{equation}
b\gamma_i^\prime\Phi^\prime(\gamma_i)+q\left(\frac{1-\Phi(\gamma_i)}{f(\gamma_i)}\right)^2\left[f(\gamma_i)-p_i\gamma_i^\prime f^\prime(\gamma_i)\right]=0,
\label{9}
\end{equation}
where $\gamma_i^\prime = \displaystyle{\frac{\dd \gamma_i}{\dd p_i}} = \displaystyle{\frac{\gamma_i}{p_i}}$, $f^\prime = \displaystyle{\frac{\dd f}{\dd \gamma_i} }$ and $\Phi^\prime = \displaystyle{\frac{\dd \Phi}{\dd \gamma_i}}$. \\

Authors in \cite{VarmaICC2012,VarmaTVT2014} proved that such equation has a unique best response. In the game $\mathcal{G}$, this best response defines the NE and is denoted as $\mathbf{p}^{\ast}=(p_1^{\ast},p_2^{\ast},\ldots,p_N^{\ast})$. However, the NE solution is not always Pareto-efficient for many scenarios. We highlight in Fig. \ref{fig:achievableregion} that the NE is not on the Pareto frontier (the outer boundary of the achievable utilities region). Therefore, we are motivated to design a more efficient solution than the NE. For this, as a first step we investigate the NB solution.

\section{Nash bargaining solution}
\label{sec:NashBargainingSolution}

Due to the inefficiency of the NE, a Pareto-efficient solution can be achieved by introducing the cooperation between the players. The resulting solution is called NB solution whose determination requires two elements \cite{Abidi2007}:
\begin{itemize}
\item the region of achievable utilities formed by the set of the feasible utilities of all the players should be compact and convex \cite{Hossain2009};
\item the threat point is defined by the NE of the one-shot game \cite{Larsson2008}.
\end{itemize}
\subsection{Compactness and convexity of the achievable utilities region}
\label{sec:CompactnessAndConvexityOfTheAchievableUtilitiesRegion}

\begin{figure}[!h]
\centering
\includegraphics[scale=0.6]{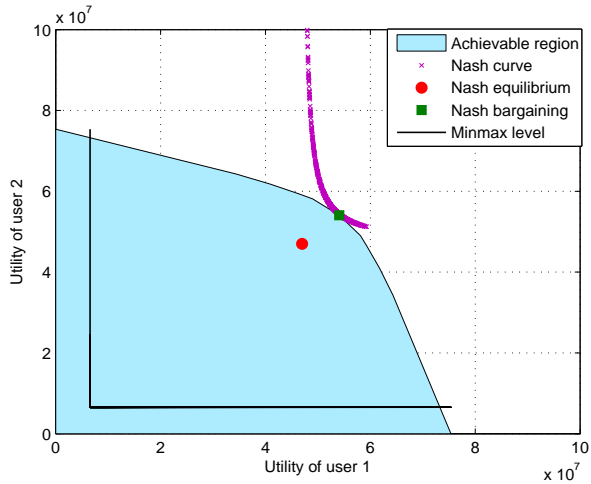}
\caption{Pareto-efficiency of the NB solution vs the NE.}
\label{fig:achievableregion}
\end{figure}

We denote $\mathcal{R}$ the achievable utilities region defined as follows:
\begin{equation}
\mathcal{R} = \{\left(u_{1},u_{2},\ldots,u_N \right)|\left(p_{1},p_{2},\ldots,p_N \right)\in \left[0,P_i^{\max}\right]^N\}.
\label{19}
\end{equation}
As the strategies sets $\mathcal{S}_1,\ldots,\mathcal{S}_N$ are compact since $\mathcal{S}_i = [0,P_i^{\max}]$ and the utility function $u_{i}$ is continuous, the region $\mathcal{R}$ is compact for a given channel configuration \cite{Larsson2008}. Since it is generally not convex, time-sharing has been a solution to convexify it. In order to illustrate the main idea of this technique applied to our problem, let us consider a system of 2 users \cite{Larsson2008}. During a time fraction $\tau$, the users use the powers $(p_1,p_2)$ to have utilities $(u_{1},u_{2})$. During a time fraction $(1-\tau)$, they use another combination of powers $(p_1',p_2')$ to have $(u_{1}',u_{2}')$ \cite{MhiriNov2012,Larsson2008}. Thus, the new achievable utilities region (for the 2-users system) is:
\begin{equation}
\begin{aligned}
\bar{\mathcal{R}} = &\{\left(\tau u_{1}+(1-\tau)u_{1}',\tau u_{2}+(1-\tau)u_{2}'\right)\\
&|0 \leq \tau \leq 1,\; \left(u_{1},u_{2}\right)\in \mathcal{R},\; \left(u_{1}',u_{2}'\right)\in \mathcal{R}\}.
\end{aligned}
\label{20}
\end{equation}
We define $\bar{\mathcal{R}}^{\ast}$ the Pareto boundary (the outer frontier) of the convex hull of $\bar{\mathcal{R}}$. Fig. \ref{fig:achievableregion} shows the convexified achievable utilities region with the NE point, the NB solution and the Nash curve (both will be defined next).
\subsection{Existence and uniqueness of the NB solution}
\label{sec:ExistenceAndUniquenessOfTheNBSolution}

Let $\mathcal{R}^{NB}$ define the improvement region of utilities versus the NE and it is given by:
\begin{equation}
\mathcal{R}^{NB} = \{u_{i}\geq u_{i}^{NE} | i \in [1,\ldots,N]\}.
\end{equation}
The NB solution belongs to the region $\mathcal{R}^{NB}$. Here, in the power control game $\mathcal{G}$, there exists a unique NB solution denoted as $\mathbf{u}^{NB}=(u_1^{NB},u_2^{NB},\ldots,u_N^{NB})$ and is given by \cite{Hossain2009}:
\begin{equation}
\mathbf{u}^{NB}=\max_{\substack{u_{i} \in \mathcal{R}^{NB} \\ i \in [1,\ldots,N]}} \prod_{i=1}^{N}{(u_{i}-u_{i}^{NE})},
\label{30}
\end{equation}

Since the NE can always be reached and the achievable utility region is a compact convex set, the NB solution exists. It is unique since it verifies certain axioms: individual rationality and feasibility, independence of irrelevant alternatives, symmetry, Pareto optimality (efficiency) and independence of linear transformations \cite{Hossain2009}. 
The NB solution results from the intersection of the Pareto boundary ($\bar{\mathcal{R}}^{\ast}$) with the Nash curve whose form is $m=\prod_{i=1}^{N}(u_{i}-u_{i}^{NE})$ where $m$ is a constant chosen such that there is precisely one intersection point \cite{Larsson2008} (see Fig. \ref{fig:achievableregion}). 
Although the NB solution is Pareto-efficient, it generally requires global CSI at the transmitters due to the Nash product $(m)$ introducing all the users utilities \cite{MhiriNov2012}. For this reason, we are looking for another efficient solution through the study of the dynamic RG.

\section{Repeated games formulation}
\label{sec:RepeatedGamesFormulation}

RG consist in their standard formulation, in repeating the same static game at every time instance and the players seek to maximize their utility averaged over the whole game duration \cite{Lasaulce2009}. Repetition allows efficient equilibrium points to be implemented and which can be predicted from the one-shot static game according to the Folk theorem, which provides the set of possible Nash equilibria of the repeated game \cite{LeTreust2010,Friedman1971}. In a repeated game, certain agreements between players on a common cooperation plan and a punishment policy can be implemented to punish the deviators \cite{Lasaulce2009}. In what follows, we introduce the new OP and characterize the two RG models. 

\subsection{New OP}
\label{sec:NewOP}

The new OP consists in setting $p_i|g_{i}|^2$ to a constant $\alpha$ which is unique when maximizing the expected sum utility over all the channel states. It is given by \cite{Mhiri2013}:
\begin{equation}
\tilde{\alpha} = \arg \max_{\alpha} \mathbb{E}_{g} \left[\sum_{i=1}^{N}{u_i(\mathbf{p})}\right].
\label{14}
\end{equation} 
The power of the $i^{th}$ player is then deduced as follows:
\begin{equation}
\tilde{p}_i = \displaystyle{\frac{\tilde{\alpha}}{|g_{i}|^2}}.
\label{15}
\end{equation} 

The new OP Pareto-dominates the NE and relies on individual CSI at the transmitter. In order to implement a cooperation plan between the players, we assume in addition to the individual CSI assumption, that every player is able to know the power of the received signal at each game stage, which is denoted by \cite{LeTreust2010}:
\begin{equation}
P_y = \sigma^{2} + \sum_{\substack{i=1}}^{N} p_{i}|g_{i}|^2. 
\end{equation}
When assuming that $p_{i}|g_{i}|^2$ is set to the constant $\alpha$, the received signal power can be written as:
\begin{equation}
P_y = \alpha\frac{\gamma_i + 1}{\gamma_i}.
\end{equation}
Accordingly, each transmitter needs only its individual SINR and the constant $\alpha$ (depending only on $p_i$ and $|g_{i}|^2$) to establish the received signal power $P_y$. We assume that the data transmission is over block fading channels and that channel gains $|g_i|^2$ lie in a compact set $[\nu_i^{\min}, \nu_i^{\max}]$ \cite{LeTreust2010}. Thus, the interval to which the received signal power belongs, is $\displaystyle{\Delta = \bigg[\sigma^2, \sigma^2 + \sum_{\substack{i=1}}^{N} p_{i}\nu_i^{\max}\bigg]}$. Since the players detect a variation of the received signal power, a deviation from the cooperation plan has occurred. Indeed, when playing at the new OP, the received signal power is constant and equal to $\displaystyle{\frac{\sigma^2(\tilde{\gamma}+1)}{1-(N-1)\tilde{\gamma}}}$. Consequently, when any player deviates from the new OP, the latter quantity changes and the deviation is then detected \cite{LeTreust2010}.

\subsection{Repeated games characterization}
\label{sec:RepeatedGamesCharacterization}

A RG is a long-term interaction game where players react to past experience by taking into account what happened in all previous stages and make decisions about their future choices \cite{Hart2006,Sorin1992}. The resulting payoff is an average over all the stage payoffs. 
We denote by $t$, the game stage which corresponds to the instant in which all players choose their actions. Accordingly, a profile of actions can be defined for all players as $\mathbf{p}(t)=(p_1(t),\allowbreak p_2(t),\ldots,p_N(t))$. A history $\mathbf{h}(t)$ of player $i$ at time $t$ is the pair of vectors $(P_{y,t},p_{i,t}) = (P_y(1),P_y(2), \allowbreak  \ldots,P_y(t-1),p_i(1),p_i(2),\ldots,p_i(t-1))$ and which lies in the set $\mathcal{H}_t = (\Delta^{t-1},\mathcal{P}_i^{t-1})$ with $\mathcal{P}_i = [0,P_i^{\max}] = [0,P^{\max}]$ (as all the users have the same maximum power) \cite{LeTreust2010}. Histories are fundamental in RG as they allow players to coordinate their behavior at each stage so that previous histories are known by all the players \cite{Sorin1992}. We denote $\delta_{i,t}$ the pure strategy of the $i^{th}$ player. It defines the action to select after each history \cite{LeTreust2010,Sorin1992}: 
\begin{equation}
\delta_{i,t} = \left | \begin{array}{rcl}
\mathcal{H}_t & \rightarrow & [0, P_i^{\max}]\\
\mathbf{h}(t) & \mapsto & p_i(t)
\end{array} \right.
\label{8}
\end{equation}
In RG literature, there are two important models \cite{Lasaulce2011}:
\begin{itemize}
\item the finite RG where the number of stages of the game (denoted as $T\geq 1$) during which the players interact is finite; 
\item the discounted RG where the discount factor (denoted as $\lambda \in ]0,1[$) is seen as the stopping probability at each stage. 
\end{itemize}
The utility function of each player results from averaging over the instantaneous utilities over all the game stages in the FRG while it is a geometric average of the instantaneous utilities during the game stages in the DRG \cite{LeTreust2010,Sorin1992,Aumann1994}. We denote $\boldsymbol{\delta}=(\delta_1,\delta_2,\ldots,\delta_N)$ the joint strategy of all players.

\begin{definition}
\textit{A joint strategy $\boldsymbol{\delta}$ satisfies the equilibrium condition for the repeated game defined by $\bigl(\mathcal{N}, (\mathcal{S}_i)_{i \in \mathcal{N}}, (v_i)_{i \in \mathcal{N}} \bigr)$ if $\forall i \in \mathcal{N}$, $\forall \delta'_i$, $v_i(\boldsymbol{\delta}) \geq v_i(\delta'_i, \boldsymbol{\delta}_{-i})$ with $v_i = v_i^T$ for the FRG or $v_i = v_i^{\lambda}$ for the DRG such that:}
\begin{align}
v_i^{T}(\boldsymbol{\delta})=   & \frac{1}{T}\sum_{t=1}^{T}u_i(\mathbf{p}(t)) & \mbox{  for the FRG}\\
v_i^{\lambda}(\boldsymbol{\delta})= &\sum_{t=1}^{+\infty}\lambda(1-\lambda)^{t-1}u_i(\mathbf{p}(t)) & \mbox{  for the DRG}
\end{align}
\end{definition}

In RG with complete information and full monitoring, the Folk theorem characterizes the set of possible equilibrium utilities. It ensures that the set of NE in a RG is precisely the set of feasible and individually rational outcomes of the one-shot game \cite{Hart2006,Sorin1992}. A cooperation/punishment plan is established between the players before playing \cite{LeTreust2010}. The players cooperate by always transmitting at the new OP with powers $\tilde{p}_i$. When the power of the received signal changes, a deviation is then detected and the players punish the deviator by transmitting with their maximum transmit power $P_i^{\max}$ in the FRG and by playing at the one-shot game in the DRG. In what follows, we give the equilibrium solution of each repeated game model and mention the corresponding algorithm \cite{Xu2012,Li2010,Song2011}. It is important to note that in contrast with iterative algorithms (e.g., iterative water-filling type algorithms), there is no convergence problem in repeated games (FRG and DRG). Indeed, the transmitters implement an equilibrium strategy (referred to as the operating point) at every stage of the repeated game.

\subsubsection{Finite RG}
\label{sec:FiniteRG}

The FRG is characterized by the minimum number of stages ($T_{\min}$). If the number of stages in the game $T$ verifies $T>T_{\min}$, a more efficient equilibrium point can be reached. However, if it is less than $T_{\min}$, the NE is then played. Assuming that channel gains $|g_i|^2$ lie in a compact set $[\nu_i^{\min}, \nu_i^{\max}]$ \cite{LeTreust2010}, we have the following proposition \cite{Mhiri2013}:

\begin{proposition}[FRG equilibrium]:
When supposing the following condition is met: $T\geq T_{\min}$ with:
\begin{equation}
T_{\min} = \left\lceil \frac{\Theta}{\Lambda-\Omega}\right\rceil,
\label{17}
\end{equation}
such that:
\begin{equation*}
\begin{array}{lcl}
\Theta & = & \displaystyle{\frac{A\nu_i^{\max}}{{b\nu_i^{\min}+\bar{\gamma}_i\sigma^2B}}- \frac{G\nu_i^{\max}}{{b\nu_i^{\min}+\tilde{\alpha}H}}}\\

\Lambda & = & \displaystyle{\frac{E\nu_i^{\min}}{{b\nu_i^{\max}+\gamma_i^{\ast}\left(\sigma^2+\sum_{j \neq i}{p_j^{\ast} \nu_i^{\max}}\right)F}}}\\

\Omega & = & \displaystyle{\frac{C\nu_i^{\min}}{{b\nu_i^{\max}+\widehat{\gamma}_i\left(\sigma^2+\sum_{j \neq i}{p_j^{\max} \nu_i^{\max}}\right)D}}}\\
\end{array}
\end{equation*}
Then, the NE corresponding to the $T$-stage FRG is given by the following action plan for any $(T,T_{\min})$ and $\forall t \geq 1$:
\begin{equation}
\delta_{i,t} : \left | \begin{array}{ll}
\tilde{p}_i & \mbox{ for }t \in \{1,2,\ldots,T-T_{\min}\}\\
p_i^{\ast} & \mbox{ for }t \in \{T-T_{\min}+1,\ldots,T\}\\
P_i^{\max} & \mbox{ for any deviation detection}
\end{array} \right.
\label{18}
\end{equation}
\end{proposition} 

The quantities $A$, $B$, $C$, $D$, $E$, $F$, $G$ and $H$ are defined in App. A and $\gamma_i^{\ast}$ is the SINR at the NE while $\bar{\gamma}_i$ and $\widehat{\gamma_i}$ are the SINRs related to the maximal utility and the utility min-max respectively (the proof of this proposition is detailed in \cite{Mhiri2013}). The corresponding algorithm is as follows.
\vspace{0.3cm}\\
\begin{tabular*}{1\textwidth}{l}
  \hline
  \makebox[\linewidth][l]{\textbf{Algorithm $1$:} FRG Algorithm}\\
  \hline
$1)$  Each user transmits at the new OP with power $\tilde{p}_i$ during the first phase \\
\hspace{0.28cm} of the game $t \in \{1,2,\ldots,T-T_{\min}\}$.\\
$2)$ In the second phase $t \in \{T-T_{\min}+1,\ldots,T\}$, each user plays the NE. \\
\hspace{0.28cm} As the FRG has a finite number of stages, this phase ensures the \\
\hspace{0.28cm} punishment of the deviator for two reasons \cite{LeTreust2010}:\\

\hspace{0.58cm} $\diamond$ if it deviates at the last stage, it cannot therefore be punished;\\
\hspace{0.58cm} $\diamond$ if it deviates earlier, the punishment can be not sufficiently severe.\\

$3)$ The power of the received signal is assumed to be constant during the \\
\hspace{0.28cm} first phase. When it changes, a deviation is then detected.\\
$4)$ The deviator is punished by other transmitters by playing at their \\
\hspace{0.28cm} maximum transmit power $P_i^{\max}$.\\
\hline
\end{tabular*}

\subsubsection{Discounted RG}
\label{sec:DiscountedRG}

In the DRG, the probability that the game stops at stage $t$ is $\lambda(1-\lambda)^{t-1}$ with $\lambda \in ]0,1[$ defines the discount factor \cite{Lasaulce2011}. Accordingly, we can express the analytic form of the maximum discount factor in a DRG when assuming that channel gains $|g_i|^2$ lie in a compact set $[\nu_i^{\min}, \nu_i^{\max}]$ \cite{LeTreust2010}.

\begin{proposition}[DRG equilibrium]:
When assuming the following condition is met: 
\begin{equation}
{\lambda \leq \frac{\Psi}{\Gamma+\Psi}},
\label{17}
\end{equation}
with:
\begin{equation*}
\begin{array}{lcl}
\Gamma & = & \displaystyle{\frac{A\nu_i^{\max}}{{b\nu_i^{\min}+\bar{\gamma}_i\sigma^2B}} - \frac{G\nu_i^{\max}}{{b\nu_i^{\min}+\tilde{\alpha}H}}}\\
\Psi & = & \displaystyle{\frac{G\nu_i^{\min}}{{b\nu_i^{\max}+\tilde{\alpha}H}}-\frac{E\nu_i^{\min}}{{b\nu_i^{\max}+\gamma_i^{\ast}\left(\sigma^2+\sum_{j \neq i}{p_j^{\ast} \nu_i^{\max}}\right)F}}}
\end{array}
\end{equation*}
Then, the NE corresponding to the DRG is given by the following action plan $\forall t \geq 1$:
\begin{equation}
\delta_{i,t} = \left | \begin{array}{ll}
\tilde{p}_i & \mbox{ when all other players play }\tilde{\mathbf{p}}_{-i}\\
p_i^{\ast} & \mbox{ else }
\end{array} \right.
\label{19}
\end{equation}
\end{proposition} 
 For the proof, see App. A. The corresponding algorithm is as follows.
\vspace{0.3cm}\\
\begin{tabular*}{1\textwidth}{l}
  \hline
  \makebox[\linewidth][l]{\textbf{Algorithm $2$:} DRG Algorithm}\\
  \hline
$1)$ Each user transmits at the new OP with power $\tilde{p}_i$. \\

$2)$ When the power of the received signal changes, a deviation is detected.\\

$3)$ The other transmitters punish the deviator by transmitting at the one-shot \\
\hspace{0.28cm} game with power $p_i^{\ast}$.\\
\hline
\end{tabular*}

\section{Numerical results}
\label{sec:NumericalResults}

In this section, we consider the efficiency function $f(x)=e^{-c/x}$ with $c=2^{\frac{R}{R_0}}-1$. It has be proven in \cite{Belmega2009,Belmega2011} that such a function is sigmoidal as it is convex on the open interval $(0,c/2]$ and concave on $(c/2,+\infty)$. The throughput $R$ and the used bandwidth $R_0$ are equal to 1 Mbps and 1 MHz respectively. The maximum power $P^{\max}$ is set to 0.1 Watt while the noise variance is set to $10^{-3}$ Watt. The buffer size $K$, the packet arrival rate $q$ and the consumed power $b$ are fixed to 10, 0.5 and $5\times 10^{-3}$ Watt respectively. We consider Rayleigh fading channels and a spreading factor $L$ introducing an interference processing ($1/L$) in the interference term of the SINR. 

\begin{figure}[!h]
\centering
\includegraphics[scale=0.6]{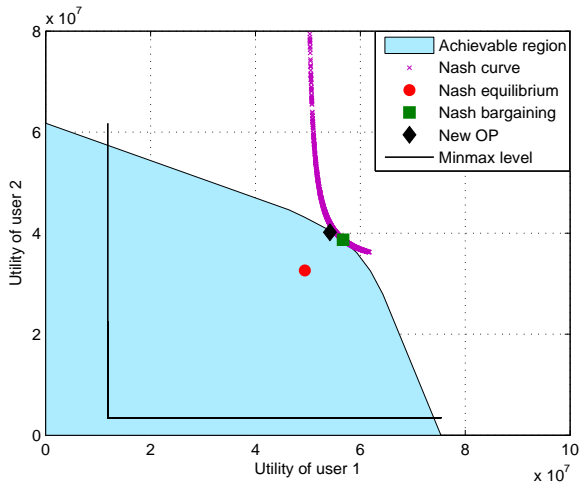}
\caption{Pareto-dominance of the new OP and the NB solution vs the NE ($L=2$).}
\label{REG2}
\end{figure}

In Fig. \ref{REG2}, we present the achievable utility region, the new OP, the NE and the NB solution. We stress that the new OP and the NB solution dominate both the NE in the sense of Pareto. The region between the Pareto frontier and the min-max level is the possible set of equilibrium utilities of the RG according to the Folk theorem.

\begin{figure}[!h]
\centering
\includegraphics[scale=0.6]{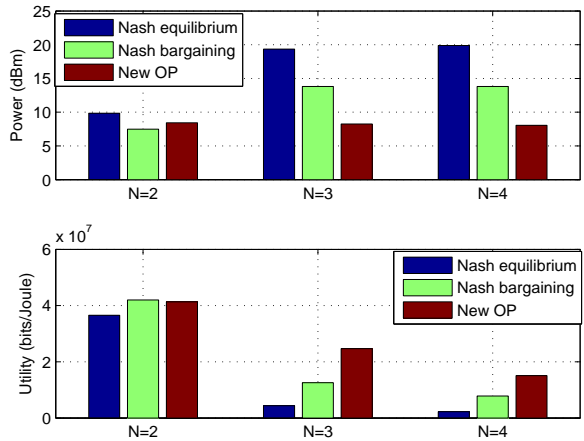}
\caption{Better performances in terms of power and utility with the new OP for different number of users $N$.}
\label{REG4}
\end{figure}

In order to study the efficiency of the new OP versus the NB solution and the NE, we are interested in comparing powers and utilities of the three equilibria by averaging over channel gains for different scenarios (different number of users $N$ in the system). In Fig. \ref{REG4}, we plot the power and the utility that a user (in a system of $N$ users) can reach for each equilibrium. Thus, we highlight that the new OP and the NB solution have better performances than the NE as they Pareto-dominate it. When $N=2$, we notice that the new OP and the NB solution are more efficient than the NE. It is clear that the NB solution requires less power and provides higher utility compared to the new OP, but it is important to stress that values, in terms of powers and utilities, are slightly different for both equilibria (new OP and NB solution). When $N>2$, we highlight that lower powers are provided with the new OP which leads also to higher values of the utilities. Thus, we notice that the new OP gives better performances than the NE and the NB solution. Therefore, the new OP contributes not only to improve the system performances better than the NE for any given scenario but also enables important gains in terms of powers and utilities when compared to the NB solution for a system with a large number of users ($N>2$).

We are interested in studying the performances of the social welfare ($\sum_i u_i$) according to the FRG versus the NE in a multi-users system. The corresponding expression is given by:
\begin{equation}
\frac{w_{FRG}}{w_{NE}}=\frac{\sum_{i=1}^{N}( \sum_{t=1}^{T-T_{\min}}\tilde{u}_i{(\mathbf{p}(t))}  + \sum_{t=T-T_{\min}+1}^{T}u^{\ast}_i{(\mathbf{p}(t))})}{\sum_{i=1}^{N}\sum_{t=1}^{T}u^{\ast}_i{(\mathbf{p}(t))}}.
\label{wfeqn}
\end{equation}

\begin{figure}[htbp]
\centering
\includegraphics[scale=0.6]{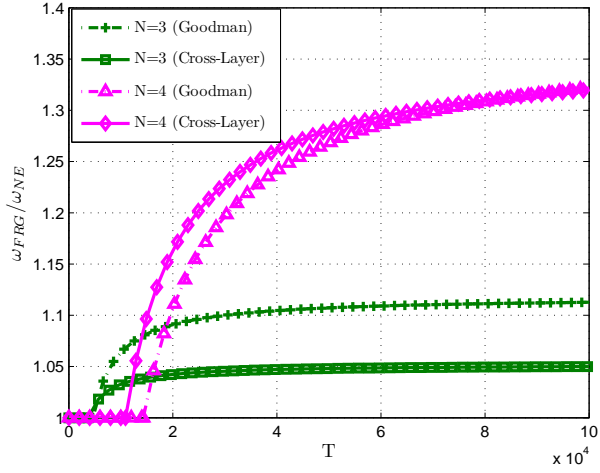}
\caption{Improvement of the social welfare in FRG vs the NE as a function of the number of stages of the game $T$ ($L=5$).}
\label{REG5}
\end{figure}

In Fig. \ref{REG5}, we present the ratio of the social welfare corresponding to the FRG ($\omega_{FRG}$) vs the NE social welfare ($\omega_{NE}$). We proceed by averaging over channel gains lying in a compact set such that $10\log_{10}(\nu^{\max}/\nu^{\min})=20$. We highlight that the social welfare of the FRG reaches higher values than the NE ($\omega_{FRG}>\omega_{NE}$). In addition, we notice that the social welfare ratio increases with the number of users for both models (Goodman and cross-layer). The minimum number of stages $T_{\min}$ according to the cross-layer model is much lower compared to the one related to the Goodman model. To illustrate this, when $N=3$, $T_{\min}$ for the Goodman model is equal to 4600 while it is 3700 for the cross-layer model. This difference becomes considerable with the increase of the number of users. Indeed, when $N=4$, the minimum number of stages for the Goodman EE is 14300 while it is equal to 10900 for the cross-layer approach. 


\begin{figure}[htbp]
\centering
\includegraphics[scale=0.6]{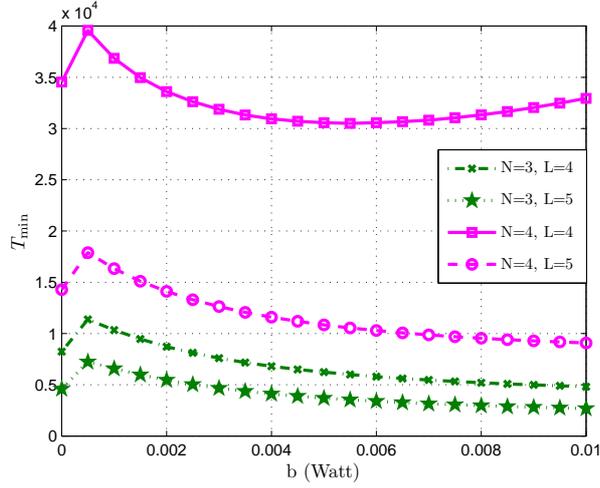}
\caption{Existence of $T_{\min}$ for the cross-layer model ($b\neq 0$) lower than $T_{\min}$ of Goodman model ($b=0$).}
\label{REG6}
\end{figure}

\begin{figure}[htbp]
\centering
\includegraphics[scale=0.6]{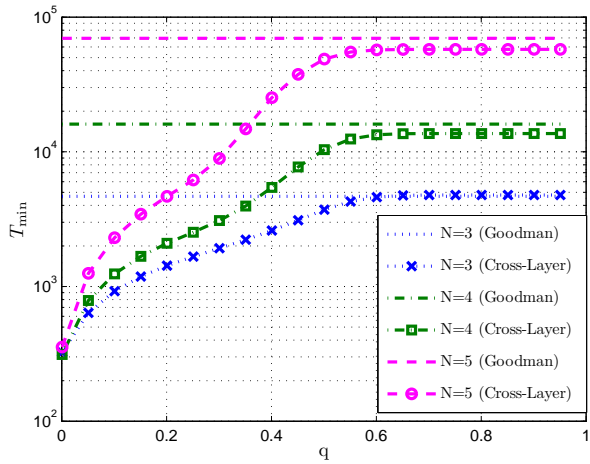}
\caption{Lower values of $T_{\min}$ of the cross-layer model when comparing to Goodman model ($L=5$).}
\label{REG11}
\end{figure}

We are interested in plotting the minimum number of stages as a function of the consumed power $b$ and the packet arrival rate $q$ according to both EE models. Results, obtained by averaging over channel realizations, are drawn in figures \ref{REG6} and \ref{REG11}. According to Fig. \ref{REG6}, we stress that $T_{\min}$ increases with the number of users while it decreases with the spreading factor. It is clear that for any values of $N$ and $L$, it exists a consumed power $b\neq 0$ for which $T_{\min}$ is less than $T_{\min}$ when $b=0$. Thus, a good choice of the fixed consumed power leads to a lower minimum number of stages for the cross-layer model compared to the Goodman model. 


In Fig. \ref{REG11}, we highlight that the minimum number of stages is an increasing function of the packet arrival rate $q$ according to the cross-layer model while it is a constant function for the Goodman model since the latter does not take into account the packet arrival process. One can confirm that the minimum number of stages is an increase function of the number of users as deduced previously. Simulations show that it exists a packet arrival rate $q_0$ before which $T_{\min}$ of the cross-layer model is much lower than $T_{\min}$ of the Goodman model for different number of users. Simulations show that $q_0 \approx 0.6$ and for $q\geq q_0$, $T_{\min}$ of the cross-layer model converges to $T_{\min}$ corresponding to the Goodman model. It is important to highlight that when $N=3$ and $q\geq q_0$, $T_{\min}$ of the cross-layer model takes higher values than $T_{\min}$ corresponding to the Goodman model but values are quite similar. With the increase of the number of users, the difference between the minimum number of stages for both models becomes noticeable. According to figures \ref{REG6} and \ref{REG11}, one can conclude that the cross-layer model can be exploited for short games.


\begin{figure}[htbp]
\centering
\includegraphics[scale=0.6]{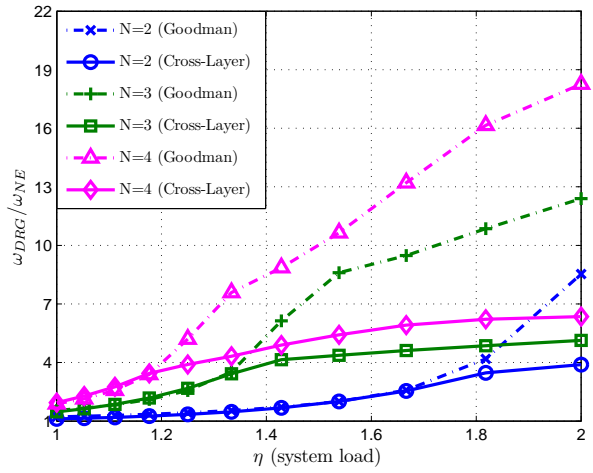}
\caption{Improvement of the social welfare in DRG vs the NE for Goodman and cross-layer models as a function of the spectral efficiency $\eta$ for different number of users $N$.}
\label{REG7}
\end{figure}

For the DRG model, we plot in a first step the improvement of the social welfare ($\omega_{DRG}$) versus the one-shot game ($\omega_{NE}$) for Goodman  and cross-layer models ($b=0$ and $b=5\times 10^{-3}$ respectively) as a function of the spectral efficiency $\eta=N/L$. We simulated our algorithm by averaging over channel gains for different number of users. Results are given in Fig. \ref{REG7}. It is important to highlight that the DRG social welfare reaches higher values than the NE social welfare ($\omega_{DRG}>\omega_{NE}$). For low values of the spectral efficiency, the social welfare ratio is quite similar for both models while the difference becomes noticeable when the spectral efficiency takes higher values. The social welfare ratio increases with the number of users for both EE models. For each model, when $N$ takes high values, the social welfare ratios become closer (for the cross-layer model, the curves corresponding to $N=3$ and $N=4$ are closer than with the curve of $N=2$). 


\begin{figure}[htbp]
\centering
\includegraphics[scale=0.6]{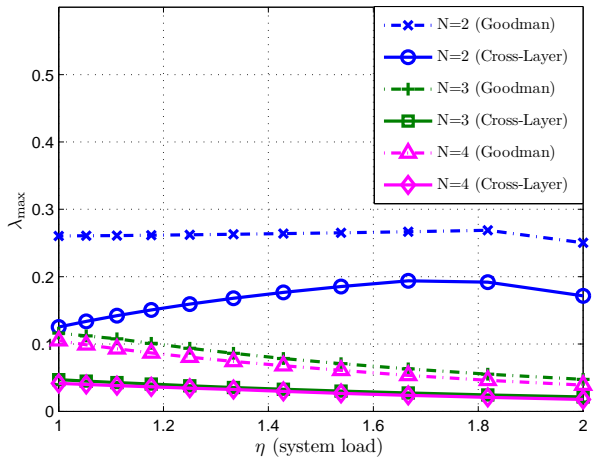}
\caption{Variation of $\lambda_{\max}$ for Goodman and cross-layer models as a function of the spectral efficiency $\eta$ with different number of users $N$.}
\label{REG8}
\end{figure}

\begin{figure}[htbp]
\centering
\includegraphics[scale=0.6]{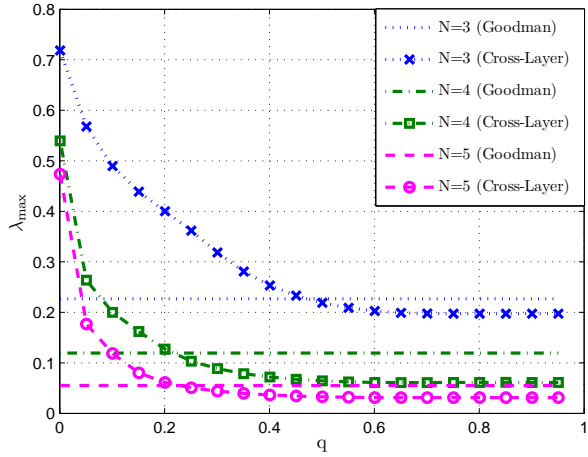}
\caption{Variation of $\lambda_{\max}$ as a function of the packet arrival rate $q$ ($L=2$).}
\label{REG10}
\end{figure}

For this reason, we studied the variation of $\lambda_{\max}$ as a function of $\eta$ and $q$ for both EE models and for different number of users. Results are given in figures \ref{REG8} and \ref{REG10}. According to Fig. \ref{REG8}, we deduce how $\lambda_{\max}$ decreases with the number of users for both EE models. In addition, we stress that the values reached by $\lambda_{\max}$ becomes closer when $N$ takes higher values. This can explain Fig. \ref{REG7}. 


The study of the variation of $\lambda_{\max}$ versus the packet arrival rate $q$ (in Fig. \ref{REG10}) shows that the maximum discount factor $\lambda_{\max}$ decreases with the number of users and with the packet arrival rate $q$ as well. Simulations show that it exists a packet arrival rate $q_1$ before which the $\lambda_{\max}$ corresponding to the cross-layer model takes higher values than the maximum discount factor of the Goodman model for different number of users. We notice that starting from $q_1$, the maximum discount factor of the cross-layer model converges to $\lambda_{\max}$ corresponding to the Goodman model. 


\begin{figure}[htbp]
\centering
\includegraphics[scale=0.6]{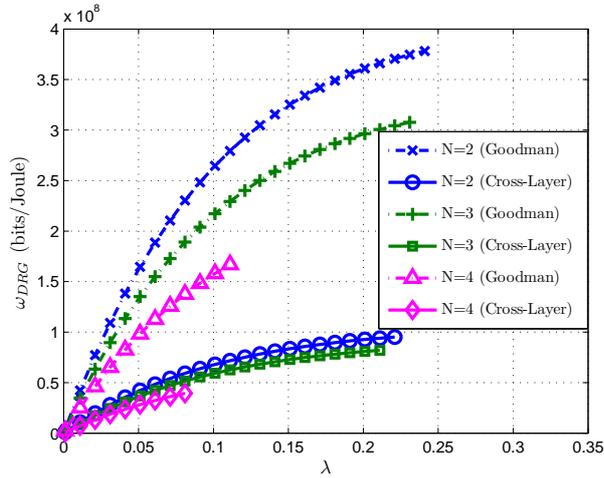}
\caption{Increase of $\omega_{DRG}$ as a function of $\lambda$ ($L=2$).}
\label{REG9}
\end{figure}

In a second step, we plotted in Fig. \ref{REG9} the variation of the DRG social welfare as a function of $\lambda \leq \lambda_{\max}$. We notice that $\omega_{DRG}$ is an increase function of $\lambda$. Thus, when $\lambda=\lambda_{\max}$, $\omega_{DRG}$ reaches highest value. However, we stress that $\omega_{DRG}$ decreases with the number of users especially for the Goodman model while it is quite similar for the cross-layer model. This confirms that the proposed new OP is still quite efficient and can be utilized for games with high number of users.  


Finally, we plot for both RG models (FRG and DRG) the social welfare when using the cross-layer approach against the constant power $b$ for two different values of the packet arrival rate $q$ ($0.5$ and $0.7$). The considered system is composed of $2$ users and the spreading factor $L$ is fixed to $4$. The idea consists in studying the efficiency of the cross-layer approach regarding the Goodman power control algorithm. Accordingly, for each packet arrival rate, we plot the social welfare with the cross-layer approach (powers at the equilibrium are determined normally according to $q$) and the social welfare with the cross-layer power control but when powers at the equilibrium are determined by the Goodman algorithm $(p[q\rightarrow1])$. Indeed, the packet arrival rate is assumed constant in the Goodman model and equal to 1 (packets arrive with probability $q=1$). For both RG models, we stress that the cross-layer power control approach outperforms the Goodman algorithm for both values of the packet arrival rate $q$. Important (relative) gains are reached. To illustrate this, for $q=0.5$ and $b=0.045$ Watt the relative gain is higher than $50\%$ in the FRG and the DRG as well. Therefore, we conclude that the OP with the cross-layer approach provides better performances and is more efficient than the OP with the Goodman power control approach.

\begin{figure}[!h]
\centering
\includegraphics[scale=0.6]{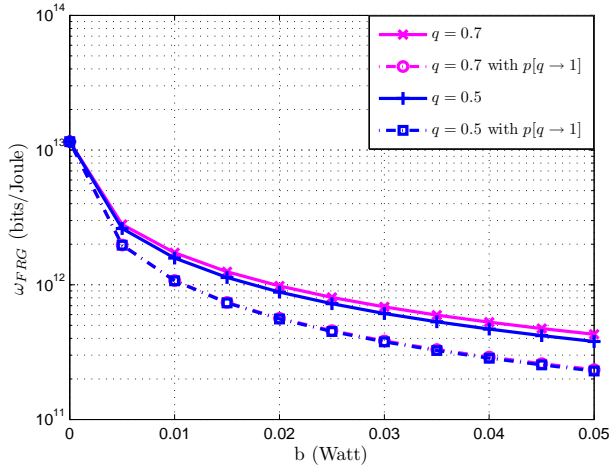}
\caption{Variations of the FRG social welfare against $b$ for $q=0.5$ and $q=0.7$: the cross-layer power control approach outperforms the Goodman algorithm. }
\label{FRG}
\end{figure}

\begin{figure}[!h]
\centering
\includegraphics[scale=0.6]{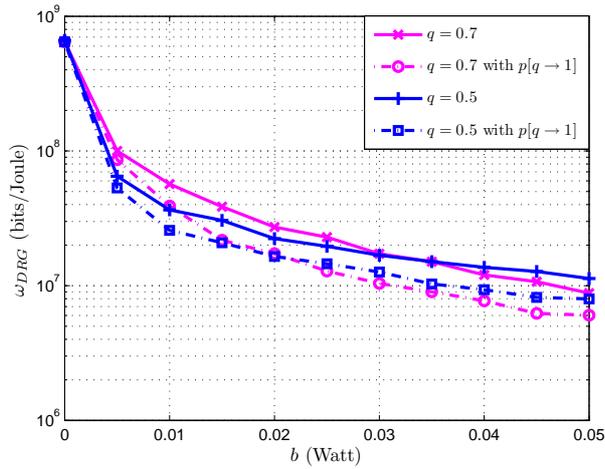}
\caption{Plotting the DRG social welfare against $b$ for $q=0.5$ and $q=0.7$: the cross-layer approach improves the power control when compared to the Goodman algorithm.}
\label{DRG}
\end{figure}

\section{Conclusion}
\label{sec:Conclusion}

In this paper, we have investigated RG for distributed power control in a MAC system. As the NE is not always energy-efficient, the NB solution might be a possible efficient solution since it is Pareto-efficient. However, the latter, in general, requires global CSI at each transmitter node. Thus, we were motivated to investigate using the repeated game formulation and develop a new OP, that simultaneously is both more efficient than the NE and achievable with only individual CSI being required at the transmitter. Also, we consider a new EE metric taking into account the presence of a queue at the transmitter with an arbitrary packet arrivals. 

Cooperation plans are proposed where the new OP is considered and closed-form expressions of the minimum number of stages for the FRG and the maximum discount factor for the DRG have been established. The study of the social welfare (sum of utilities of all the users) shows that considerable gains are reached compared to the NE (for the FRG and DRG). Moreover, our model proves that even with a high number of users, the FRG can always be played with a minimum number of stages shorter than when using the Goodman model. In addition, the social welfare in the DRG decreases slightly with the number of users with the cross-layer approach while it decreases considerably with the Goodman model. Finally, the comparison of the cross-layer algorithm versus the Goodman algorithm, shows that in real systems with random packet arrivals, the cross-layer power control algorithm outperforms the Goodman algorithm. Thus, the new OP with the cross-layer approach is more efficient. An interesting extension to this work would be to consider the interference channel instead of the MAC channel and generalize the framework applied here. Another possible extension would be to consider the multi-carrier case and the resulting repeated game.

\begin{appendices}
\section{(Proof of $\lambda_{max}$)}
\label{sec:Proof}
\subsection{Determination of the maximal utility}
\label{sec:DeterminationOfTheMaximalUtility}

Let us determine the maximal utility that a player can get and which is denoted as follows: 
\begin{equation}
\begin{array}{lcl}
\displaystyle{\bar{u}_i }&= &\displaystyle{\max_{\mathbf{p}_{-i}}\max_{p_i}u_i({p_i,\mathbf{p}_{-i}})}.\\
\end{array}
\label{A1}
\end{equation}

We denote $\dot{p}_i$ the power maximizing the utility function $u_i$ and which is the solution of the following equation:
\begin{equation}
\displaystyle{b\frac{\gamma_i}{p_i}\Phi^\prime(\gamma_i)+q\left(\frac{1-\Phi(\gamma_i)}{f(\gamma_i)}\right)^2 \left[f(\gamma_i)-\gamma_i f^\prime(\gamma_i)\right]=0},
\label{A3}
\end{equation}
with $\displaystyle{\gamma_i^\prime = {\frac{\dd \gamma_i}{\dd p_i}} = {\frac{\gamma_i}{p_i}}}$, $\displaystyle{f^\prime = {\frac{\dd f}{\dd \gamma_i} }}$ and $\displaystyle{\Phi^\prime = {\frac{\dd \Phi}{\dd \gamma_i}}}$. Therefore, the expression of the maximum utility function writes as:
\begin{equation}
\dot{u}_{i}({\dot{p}_i,\mathbf{p}_{-i}}) = \displaystyle{\frac{Rq(1-\phi(\dot{\gamma}_{i}))}{{b+\displaystyle{\frac{\dot{p}_{i}q(1-\phi(\dot{\gamma}_{i}))}{f(\dot{\gamma}_{i})}}}}},
\label{A4}
\end{equation}
with: 
\begin{equation}
\displaystyle{\dot{\gamma}_{i} = \frac{\dot{p}_i|g_i|^2}{\sigma^2+\sum_{j \neq i}{p_j |g_j|^2}}}.
\label{A5}
\end{equation}

We have to study then the behavior of $\dot{u}_{i}({\dot{p}_i,\mathbf{p}_{-i}})$ regarding $p_j$ for $j\neq i$ and then we determine the sign of $\displaystyle{\frac{\partial \dot{u}_{i}({\dot{p}_i,\mathbf{p}_{-i}})}{\partial p_j}}$ which is given by:
\begin{equation}
\displaystyle{\frac{\partial \dot{u}_{i}({\dot{p}_i,\mathbf{p}_{-i}})}{\partial p_j}} = \displaystyle{Rq\frac{\displaystyle{\frac{-b\partial \phi(\dot{\gamma}_{i})}{\partial p_j}+ \dot{p}_i q   \bigg(\frac{(1-\phi(\dot{\gamma}_{i}))}{f(\dot{\gamma}_{i})}\bigg)^2\frac{\partial f(\dot{\gamma}_{i})}{{\partial p_j}}}}{\displaystyle{\bigg(b+\frac{\dot{p}_i q(1-\phi(\dot{\gamma}_{i}))}{f(\dot{\gamma}_{i})}\bigg)^2}}}.
\label{5}
\end{equation}

We are interested to study the sign of the numerator:
\begin{multline}
\frac{-b\partial \phi(\dot{\gamma}_{i})}{\partial p_j}+ \dot{p}_i q \bigg(\frac{(1-\phi(\dot{\gamma}_{i}))}{f(\dot{\gamma}_{i})}\bigg)^2\frac{\partial f(\dot{\gamma}_{i})}{\partial p_j} = \\
\bigg(\frac{-b\partial \phi(\dot{\gamma}_{i})}{\partial \dot{\gamma}_{i}}+ \dot{p}_i q \bigg(\frac{(1-\phi(\dot{\gamma}_{i}))}{f(\dot{\gamma}_{i})}\bigg)^2\frac{\partial f(\dot{\gamma}_{i})}{\partial \dot{\gamma}_{i}}\bigg)\frac{\partial \dot{\gamma}_{i}}{\partial p_j},
\label{6}
\end{multline}
with: 
\begin{equation}
\frac{\partial \dot{\gamma}_{i}}{\partial p_j} = \frac{-\dot{p}_i|g_i|^2|g_j|^2}{\bigg(\sigma^2+\sum_{j \neq i}{p_j |g_j|^2\bigg)^2}}<0.
\label{7}
\end{equation}

The next step would be to determine the sign of the expression\\
$\displaystyle {\frac{-b\partial \phi(\dot{\gamma}_{i})}{\partial \dot{\gamma}_{i}} \allowbreak + \dot{p}_i q \bigg(\frac{(1-\phi(\dot{\gamma}_{i}))}{f(\dot{\gamma}_{i})}\bigg)^2\frac{\partial f(\dot{\gamma}_{i})}{\partial \dot{\gamma}_{i}}}$. 
It is obvious that $\displaystyle {\dot{p}_i q \bigg(\frac{(1-\phi(\dot{\gamma}_{i}))}{f(\dot{\gamma}_{i})}\bigg)^2\frac{\partial f(\dot{\gamma}_{i})}{\partial \dot{\gamma}_{i}}>0}$ since $f$ is an increasing function of the SINR. Therefore, we need to determine the sign of $\displaystyle{\frac{\partial \phi(\dot{\gamma}_{i})}{\partial \dot{\gamma}_{i}}}$. We have: 
\begin{equation}
\begin{array}{lcl}
\displaystyle {\frac{\partial \phi(\gamma_i)}{\partial \gamma_i}} & = & \displaystyle {\frac{\partial ((1-f(\gamma_i))\Pi(\gamma_i))}{\partial \gamma_i}}\\
& = & \displaystyle {-\frac{\partial f(\gamma_i)}{\partial \gamma_i}\Pi(\gamma_i)+(1-f(\gamma_i))\frac{\partial \Pi(\gamma_i)}{\partial \gamma_i}}.
\end{array}
\label{8}
\end{equation}

The sign of the first term is negative while the sign of the second term is the same as $\partial \Pi(\gamma_i)/\partial \gamma_i$ since $(1-f(\gamma_i))>0$ and we have:
\begin{equation}
\displaystyle {\frac{\partial \Pi(\gamma_i)}{\partial \gamma_i}} =  \displaystyle {\frac{\partial \rho(\gamma_i)}{\partial \gamma_i}\frac{\partial \Pi(\gamma_i))}{\partial \rho}}.
\label{9}
\end{equation}
However $\rho(\gamma_i)=\displaystyle{\frac{q(1-f(\gamma_i))}{(1-q)f(\gamma_i)}}$ and then:
\begin{equation}
\displaystyle {\frac{\partial \rho(\gamma_i)}{\partial \gamma_i}} = \displaystyle {\frac{-q}{(1-q)f^2(\gamma_i)}\frac{\partial f(\gamma_i)}{\partial \gamma_i}<0}.
\label{10}
\end{equation}
As shown in \cite{VarmaICC2012}, we have: 
\begin{equation}
\Pi(\gamma_i) = \frac{\rho^K}{1+\rho+\rho^2+\ldots+\rho^K}.
\label{11}
\end{equation}
The latter quantity can be expressed as:
\begin{equation}
\frac{1}{\Pi(\gamma_i)} = 1+\frac{1}{\rho}+\frac{1}{\rho^2}+\ldots+\frac{1}{\rho^K}.
\label{12}
\end{equation}
Consequently, we have: 
\begin{equation}
\frac{\partial \Pi(\gamma_i)}{\partial \rho} = \Pi^2(\gamma_i)\bigg[\frac{1}{\rho^2}+\frac{2}{\rho^3}+\ldots+\frac{K}{\rho^{K+1}}\bigg]>0.
\label{13}
\end{equation}
Therefore, $\displaystyle {\frac{\partial \Pi(\gamma_i)}{\partial \gamma_i}<0}$ and hence $\displaystyle {\frac{\partial \phi(\gamma_i)}{\partial \gamma_i}}<0$. In particular, we have $\displaystyle{\frac{\partial \phi(\dot{\gamma}_{i})}{\partial \dot{\gamma}_{i}}}<0$. Thus, we have $\displaystyle { \bigg(\frac{-b\partial \phi(\dot{\gamma}_{i})}{\partial \dot{\gamma}_{i}}+ \dot{p}_i q \bigg(\frac{(1-\phi(\dot{\gamma}_{i}))}{f(\dot{\gamma}_{i})}\bigg)^2\frac{\partial f(\dot{\gamma}_{i})}{\partial \dot{\gamma}_{i}}\bigg)}>0$ and finally $\displaystyle{\frac{\partial \dot{u}_{i}({\dot{p}_i,\mathbf{p}_{-i}})}{\partial p_j}}<0$. We deduce then that $\dot{u}_{i}$ is a decreasing function of $p_j$. It reaches its maximum when $p_j=0$ and it is minimum when $p_j=p_j^{\max}$ (for all $j\neq i$). When substituting $p_j=0$ in the SINR expression, this allows the determination of the optimal power:
\begin{equation}
b\frac{|g_i|^2}{\sigma^2}\Phi^\prime(\gamma_i(p_i))+ q\bigg(\frac{(1-\Phi(\gamma_i(p_i)))}{f(\gamma_i(p_i))}\bigg)^2 \bigg [f(\gamma_i(p_i))-\gamma_i f^\prime(\gamma_i(p_i))\bigg]=0,
\label{17}
\end{equation}
with: $\displaystyle{\gamma_{i} = \frac{p_i|g_i|^2}{\sigma^2}}$.\\
The latter equation is a function of the SINR. We determine then the solution in terms of SINR which we denote $\bar{\gamma}_i$ and for which the optimal power is $\displaystyle{\bar{p}_i=\frac{\bar{\gamma}_i\sigma^2}{|g_i|^2}}$. This SINR exists due to the quasi-concavity of $u_i$ in $(p_i,\mathbf{p}_{-i})$ \cite{VarmaICC2012,VarmaTVT2014}. Then, we have: 
\begin{equation}
\bar{u}_i = \max_{\mathbf{p}}u_i(\mathbf{p}) = \displaystyle{\frac{Rq(1-\phi(\bar{\gamma}_i))}{\displaystyle{b+\frac{\bar{\gamma}_i\sigma^2}{|g_i|^2}\frac{q(1-\phi(\bar{\gamma}_i))}{f(\bar{\gamma}_i)}}}}.
\label{18}
\end{equation}
\subsection{Determination of $\lambda_{\max}$}
\label{sec:DeterminationOfLambdaMax}

The SINR $\tilde{\gamma}_i$ refers to the SINR when playing the new OP while $\gamma_i^{\ast}$, $\bar{\gamma}_i$ and $\widehat{\gamma_i}$ are the SINRs at the NE, at the maximal utility and at the utility min-max respectively. In order to simplify expressions, we define the following notations: 
\begin{equation*}
\begin{array}{lcl}
A & = & Rq(1-\phi(\bar{\gamma}_i))\\
B & = & \displaystyle{\frac{q(1-\phi(\bar{\gamma}_i))}{f(\bar{\gamma}_i)}}\\
C & = & Rq(1-\phi(\widehat{\gamma}_i))\\
D & = & \displaystyle{\frac{q(1-\phi(\widehat{\gamma}_i))}{f(\widehat{\gamma}_i)}}\\
E & = & Rq(1-\phi(\gamma_i^{\ast}))\\
F & = & \displaystyle{\frac{q(1-\phi(\gamma_i^{\ast}))}{f(\gamma_i^{\ast})}}\\
G & = & Rq(1-\phi(\tilde{\gamma}_i))\\
H & = & \displaystyle{\frac{q(1-\phi(\tilde{\gamma}_i))}{f(\tilde{\gamma}_i)}}
\end{array}
\label{B3}
\end{equation*}
At a stage $t$, the equilibrium condition is \cite{LeTreust2010}:
\begin{equation}
\begin{array}{c}
\lambda \bar{u}_i(\mathbf{p}(t)) + \sum_{s\geq t+1}{\lambda (1-\lambda)^{s-t}\mathbb{E}_{g}[u_i^{\ast}(\mathbf{p}(s))]} \\
\leq \lambda \tilde{u}_i(\mathbf{p}(t)) + \sum_{s\geq t+1}{\lambda (1-\lambda)^{s-t}\mathbb{E}_{g}[\tilde{u}_i(\mathbf{p}(s))]}
\end{array}
\end{equation}
Knowing that $\sum_{s\geq t+1}{(1-\lambda)^{s-t}} = (1-\lambda)/\lambda$, we have: 
\begin{equation}
\lambda \bar{u}_i + (1-\lambda)\mathbb{E}_{g}[u_i^{\ast}] \leq \lambda \tilde{u}_i+(1-\lambda)\mathbb{E}_{g}[\tilde{u}_i]
\end{equation}

\begin{equation}
\begin{array}{c}
\Longleftrightarrow \lambda {\frac{A|g_i|^2}{{b|g_i|^2+\bar{\gamma}_i\sigma^2B}}} + (1-\lambda)\mathbb{E}_{g}\left[{\frac{E|g_i|^2}{{b|g_i|^2+\gamma_i^{\ast}\left(\sigma^2+\sum_{j \neq i}{p_j^{\ast} |g_j|^2}\right)F}}}\right] \\
 \leq \lambda {\frac{G|g_i|^2}{{b|g_i|^2+\tilde{\alpha}H}}}+(1-\lambda)\mathbb{E}_{g}\left[ {\frac{G|g_i|^2}{{b|g_i|^2+\tilde{\alpha}H}}}\right]
\end{array}
\end{equation}

\begin{equation}
\begin{array}{c}
\Rightarrow \lambda \left[{\frac{A\nu_i^{\max}}{{b\nu_i^{\min}+\bar{\gamma}_i\sigma^2B}} - \frac{G\nu_i^{\max}}{{b\nu_i^{\min}+\tilde{\alpha}H}}} \right] \\
\leq (1-\lambda)\left[{\frac{G\nu_i^{\min}}{{b\nu_i^{\max}+\tilde{\alpha}H}} - \frac{E\nu_i^{\min}}{{b\nu_i^{\max}+\gamma_i^{\ast}\left(\sigma^2+\sum_{j \neq i}{p_j^{\ast} \nu_i^{\max}}\right)F}}}\right].
\end{array}
\end{equation}

Let $\Psi$ and $\Gamma$ define the following quantities : 
\begin{equation*}
\begin{array}{lcl}
\Gamma & = & \displaystyle{\frac{A\nu_i^{\max}}{{b\nu_i^{\min}+\bar{\gamma}_i\sigma^2B}} - \frac{G\nu_i^{\max}}{{b\nu_i^{\min}+\tilde{\alpha}H}}}\\
\Psi & = & \displaystyle{\frac{G\nu_i^{\min}}{{b\nu_i^{\max}+\tilde{\alpha}H}}-\frac{E\nu_i^{\min}}{{b\nu_i^{\max}+\gamma_i^{\ast}\left(\sigma^2+\sum_{j \neq i}{p_j^{\ast} \nu_i^{\max}}\right)F}}}
\end{array}
\end{equation*}

Thus: 
\begin{equation}
\displaystyle{{\lambda_{\max} = \frac{\Psi}{\Gamma+\Psi}}}.
\end{equation}

\end{appendices}

\section*{Competing interests}

The authors declare that they have no competing interests.

\end{document}